\documentclass[aps,prb,twocolumn,superscriptaddress,floatfix,longbibliography]{revtex4-2}

\usepackage{amsmath,amssymb,amsfonts,bm}
\usepackage{graphicx}
\usepackage{dcolumn}
\usepackage{hyperref}
\usepackage{xcolor} 
\usepackage{multirow}
\usepackage{makecell}

\hypersetup{
    colorlinks=true,
    linkcolor=blue,
    citecolor=blue,
    urlcolor=blue
}
\pagecolor{white}
\begin{document}

\title{Taming the 3D Wilson-Fisher Fixed Point via Nonlocal Effective Action}

\author{Hyeon Jung Kim}
\email{hkim7218@postech.ac.kr}

\author{Seung-Jong Yoo}
\email{y2ysj@postech.ac.kr}

\author{Jin Mo Bok}
\email{jinmobok@postech.ac.kr}

\author{Lemuel John Sese}
\author{Semin Park}
\affiliation{Department of Physics, POSTECH, Pohang, Gyeongbuk 37673, Korea}

\author{Ki-Seok Kim}
\email{tkfkd@postech.ac.kr}
\affiliation{Department of Physics, POSTECH, Pohang, Gyeongbuk 37673, Korea}
\affiliation{Asia Pacific Center for Theoretical Physics (APCTP), Pohang, Gyeongbuk 37673, Korea}

\date{\today}

\begin{abstract}

We present a Renormalization Group (RG) framework based on a nonlocal effective action ansatz to analyze the strong coupling dynamics of the three-dimensional relativistic $\phi^{4}$ theory. By implementing a Hubbard-Stratonovich transformation, we decouple the quartic interaction into the primary field $\phi$ and an auxiliary field $\varphi \sim \phi^2$, allowing both exponents $\Delta_{\phi}$ and $\Delta_{\varphi}$ to act as independent, unconstrained variables rather than fixed scaling dimensions. Within this nonlocal propagator framework, both the field self-energies and vertex corrections are evaluated at the one-loop order.
The resulting one-loop logarithmic derivatives determine the renormalization group flows of the couplings and the scaling exponents. 
For $d=3$ and $\epsilon\approx-0.198$, the self-consistent equations yield a representative fixed point at $\Delta_{\phi}\approx0.97714$, $\Delta_{\varphi}\approx-0.65260$, and $\Delta_{\phi^2}\approx1.04573$, corresponding to $\eta_{\phi}\approx0.04572$ and $\nu\approx0.51170$.
Relative to the high-precision conformal-bootstrap benchmarks, the deviations are approximately $0.48\%$ for $\Delta_{\phi}$, $25.97\%$ for $\Delta_{\phi^2}$, and $18.77\%$ for $\nu$, demonstrating sub-percent agreement in the fundamental-field sector while revealing substantially larger deviations in the composite and correlation-length sectors within the leading-order truncation.

\end{abstract}

\maketitle

\section{Introduction}
The Wilson--Fisher fixed point of the one-component scalar $\phi^4$ theory governs both the three-dimensional Ising universality class \cite{Wilson_Fisher_1972,Wilson_Kogut_1974} and, through the quantum-to-classical correspondence, Lorentz-invariant quantum critical points with a single real scalar order parameter and $\mathbb{Z}_2$ symmetry in $2+1$ dimensions \cite{Sachdev_2011}.
Because the fixed point is strongly coupled, determining its critical properties directly in three dimensions remains challenging.
Complementary field-theoretic approaches include expansions about four dimensions and fixed-dimension perturbation theory \cite{LeGuillou_1977,Pelissetto_Vicari_2002}, large-$N$ techniques \cite{Moshe_ZinnJustin_2003}, and functional and exact renormalization-group methods \cite{Berges_2002,Rosten_2012}.
High-precision results have also been obtained from Monte Carlo simulations \cite{Hasenbusch_2010,Ferrenberg_2018} and the conformal bootstrap \cite{ElShowk_2012,Kos_2016}.
%At fixed dimension, however, applications of functional and exact RG methods generally rely on truncations, making their accuracy approximation-dependent \cite{Berges_2002,Rosten_2012}.
%Beyond the quantitative accuracy of any particular truncation, a distinct structural question remains: can the scaling dimensions and critical exponents of this strongly coupled fixed point be determined self-consistently within a compact field-theoretic closure formulated directly at fixed $d=3$, without fixing the composite sector a priori?
%Beyond the quantitative accuracy of any particular truncation, a broader structural question remains: can a compact nonlocal fixed-point closure formulated directly in three dimensions encode the independent scaling of the fundamental and leading composite operators while maintaining consistency with the interaction that couples them?
%Beyond the quantitative accuracy of any particular truncation, a distinct structural question remains: can a compact field-theoretic closure formulated directly at $d=3$ treat the fundamental field and its leading composite as independently scaling sectors while maintaining consistency with the interaction that couples them?
%Beyond the quantitative accuracy of any particular truncation, a distinct structural question remains: can a nonlocal fixed-point closure formulated directly at $d=3$ determine the scaling of the fundamental field and its leading composite as independent sectors while maintaining consistency with the interaction that couples them?
Fixed-dimensional RG calculations nevertheless remain dependent on how the effective action is truncated.\cite{Berges_2002,Rosten_2012}.
This motivates a distinct structural question: how much of the Wilson–Fisher scaling structure can be reproduced by a minimal closure in which the fundamental field and its leading composite are allowed to scale independently, constrained only by mutual consistency with the interaction that couples them?

This question becomes concrete in the Hubbard--Stratonovich formulation: at conventional leading order in large $N$, the auxiliary field $\varphi\sim\phi^2$ inherits its momentum scaling from the $\phi$ polarization bubble \cite{Moshe_ZinnJustin_2003,Fei_Giombi_Klebanov_2014}.
We instead formulate a nonlocal effective-action ansatz by assigning independent scale-covariant kernels to $\phi$ and $\varphi$ and treating their momentum powers as variables to be determined.
The self-energy corrections to the $\phi$ and $\varphi$ kernels are then combined with the Yukawa-vertex flow to close the fixed-point equations self-consistently.
This ``unfreezing'' of the auxiliary sector turns the determination of the fixed point into a compatibility problem among the fundamental, composite, and vertex sectors.
It also provides an internal diagnostic: a candidate branch must satisfy both the interaction flow and the joint matching condition for the two nonlocal kernels.

At one loop, the coupled closure admits several nontrivial candidate branches.
We focus on a representative branch whose composite-operator dimension and corresponding correlation-length exponent lie within a few percent of high-precision Monte Carlo and conformal-bootstrap benchmarks \cite{Hasenbusch_2010,Ferrenberg_2018,Kos_2016}.
Although the level of agreement is not uniform across the fundamental, composite, and interaction sectors, this branch provides a stringent test of how much of the Wilson--Fisher scaling structure can be recovered within the minimal truncation when all three sectors are determined self-consistently.
Its overall performance thus provides an encouraging proof of concept for a self-consistent fixed-dimensional determination of the Wilson--Fisher fixed point.

\section{Effective Field Theory and Two-Variable Ansatz} \label{sec:eft_framework}
To implement this closure, we consider a nonlocal low-energy effective action for the critical three-dimensional $\phi^4$ theory in the Hubbard--Stratonovich representation.
Setting all mass terms to zero, we take the effective action to be

\begin{equation}
\begin{aligned}
\mathcal{S}_{\mathrm{EFT}}
={}&\frac{1}{2}\int_p
\Big[
(p^2)^{\Delta_\phi}\phi(p)\phi(-p)
+(p^2)^{\Delta_\varphi}\varphi(p)\varphi(-p)
\Big]
\\
&+g\int_x \varphi(x)\phi^2(x).
\end{aligned}
\label{eq:action_final}
\end{equation}

%\begin{widetext}
%\begin{equation}
%\mathcal{S}_{\text{EFT}}
%=
%\int \frac{d^dp}{(2\pi)^d}
%\left[
%\frac{1}{2}(p^2)^{\Delta_\phi}\phi(p)\phi(-p)
%+
%\frac{1}{2}(p^2)^{\Delta_\varphi}\varphi(p)\varphi(-p)
%\right]
%+
%g\int d^dx\,\varphi(x)\phi^2(x).
%\label{eq:action_final}
%\end{equation}
%\end{widetext}
The momentum-space scaling dimensions are $[\phi(p)] = -d/2 - \Delta_\phi$ and $[\varphi(p)] = -d/2 - \Delta_\varphi$, while the Yukawa coupling has the scaling dimension
\begin{equation}
[g]
=
2\Delta_\phi+\Delta_\varphi-\frac d2
\equiv \epsilon .
\label{eq:def_ep}
\end{equation}
Consequently, this nonlocal propagator ansatz replaces the conventional local derivative kinetic terms, treating the exponents $\Delta_\phi$ and $\Delta_\varphi$ as independent dynamical variables to evaluate the one-loop renormalization group equations.

In the conventional leading-order large-$N$ Hubbard--Stratonovich formulation, the polarization bubble fixes the auxiliary-field exponent to $\Delta_\varphi=-1/2$ in $d=3$ at leading order \cite{Moshe_ZinnJustin_2003,Fei_Giombi_Klebanov_2014}.
By contrast, keeping both $\Delta_\phi$ and $\Delta_\varphi$ as unfixed parameters allows the quadratic kernels to absorb the logarithmic corrections generated during the diagrammatic expansion.

The diagrammatic corrections are organized as logarithmic deformations of the two quadratic kernels and of the local Yukawa vertex. 
Based on the scaling structure of the nonlocal ansatz, the self-energies are formally expanded as
\begin{equation}
\Pi_\varphi(p)
=
(p^2)^{\Delta_\varphi}
\sum_{n\geq 1}
\tilde g^{2n}
A^{(nL)}(\Delta_\phi,\Delta_\varphi)
(p^2)^{-n\epsilon},
\label{eq:self_energy_varphi_expansion}
\end{equation}
\begin{equation}
\Sigma_\phi(p)
=
(p^2)^{\Delta_\phi}
\sum_{n\geq 1}
\tilde g^{2n}
B^{(nL)}(\Delta_\phi,\Delta_\varphi)
(p^2)^{-n\epsilon},
\label{eq:self_energy_phi_expansion}
\end{equation}
and, after factoring out the tree-level Yukawa coupling $g$, the vertex expansion takes the form
\begin{equation}
\Gamma_g(p,0)
=
\sum_{n\geq 1}
\tilde g^{2n}
C^{(nL)}(\Delta_\phi,\Delta_\varphi)
(p^2)^{-n\epsilon}.
\label{eq:vertex_expansion}
\end{equation}
Here, $A^{(nL)}$, $B^{(nL)}$, and $C^{(nL)}$ denote the corresponding $n$-loop diagrammatic coefficient functions, where $\tilde g^2\equiv g^2/(4\pi)^{3/2}$. 
The factor $(p^2)^{-n\epsilon}$ reflects the deviation of the Yukawa coupling from exact marginality at each loop order. 
Self-consistency of this formal expansion requires the condition $|\epsilon|\ll 1$, which allows $(p^2)^{-n\epsilon}$ to be treated as a controlled logarithmic correction.
\\

\section{One-Loop Evaluation and Exponent Unfreezing} \label{sec:multi_loop}

We now implement the one-loop truncation of this formal expansion. 
The diagrammatic coefficient functions $A^{(nL)}$, $B^{(nL)}$, and $C^{(nL)}$ are restricted to their leading one-loop contributions, $A^{(1L)}$, $B^{(1L)}$, and $C^{(1L)}$, while all higher-loop corrections are omitted. 
The evaluation is performed in $d$-dimensional Euclidean momentum space, treating the exponents $\Delta_\phi$ and $\Delta_\varphi$ as independent parameters.

\subsection{1-Loop Auxiliary Self-Energy}

%The auxiliary field $\varphi$ acquires its dynamics via the one-loop bubble fluctuation. 
The quadratic kernel of the auxiliary field $\varphi$ receives a momentum-dependent one-loop contribution from the $\phi$ polarization bubble.
Using the generalized propagator $G_{\phi}(k) = (k^2)^{-\Delta_{\phi}}$, the momentum-space Feynman integral is written as
\begin{equation}
\Pi_{\varphi}^{(1L)}(p) = 2g^2 \int \frac{d^dk}{(2\pi)^d} \frac{1}{(k^2)^{\Delta_{\phi}} \, ((p-k)^2)^{\Delta_{\phi}}} . \label{eq:bubble_integral}
\end{equation}
To evaluate this expression for arbitrary exponents, we employ the Feynman parameter identity. 
Applying this identity to Eq.~\eqref{eq:bubble_integral} and shifting the loop momentum to $k' = k - (1-x)p$ yields the following form:
\begin{equation}
\begin{aligned}
\Pi_{\varphi}^{(1L)}(p)
={}&
2g^2\frac{\Gamma(2\Delta_{\phi})}
{\Gamma(\Delta_{\phi})^2}
\int_0^1 dx\,
\bigl[x(1-x)\bigr]^{\Delta_{\phi}-1}
\\
&\times
\int\frac{d^dk'}{(2\pi)^d}
\frac{1}{
\bigl[k'^2+x(1-x)p^2\bigr]^{2\Delta_{\phi}}
}.
\end{aligned}
\label{eq:bubble_feynman_param}
\end{equation}
%\begin{widetext}
%\begin{equation}
%\Pi_{\varphi}^{(1L)}(p) = 2g^2\frac{\Gamma(2\Delta_{\phi})}{\Gamma(\Delta_{\phi})^2} \int_0^1 dx \, x^{\Delta_{\phi}-1}(1-x)^{\Delta_{\phi}-1} \int \frac{d^dk'}{(2\pi)^d} \frac{1}{[k'^2 + x(1-x)p^2]^{2\Delta_{\phi}}} . \label{eq:bubble_feynman_param}
%\end{equation}
%\end{widetext}
Performing the momentum integration leads to the expression:
\begin{equation}
\Pi_{\varphi}^{(1L)}(p) = \frac{2g^2}{(4\pi)^{d/2}} \frac{\Gamma(2\Delta_{\phi} - \frac{d}{2})}{\Gamma(\Delta_{\phi})^2} \int_0^1 dx \frac{x^{\Delta_{\phi}-1}(1-x)^{\Delta_{\phi}-1}}{[x(1-x)p^2]^{2\Delta_{\phi}-d/2}} . \label{eq:bubble_momentum_integrated}
\end{equation}
Evaluating the remaining Feynman parameter integral through the Euler Beta function identity determines the coefficient function $A^{(1L)}(\Delta_{\phi})$ as a ratio of Euler Gamma functions:
\begin{equation}
A^{(1L)}(\Delta_{\phi}) = \frac{2\Gamma(2\Delta_{\phi} - \frac{d}{2}) \Gamma(\frac{d}{2}-\Delta_{\phi})^2}{\Gamma(\Delta_{\phi})^2 \Gamma(d-2\Delta_{\phi})} .  \label{eq:A_final}
\end{equation}
\newline
\subsection{1-Loop ``Sunset'' Self-Energy}

The primary field $\phi$ receives its momentum-dependent renormalization from the one-loop diagram carrying one internal $\varphi$ line and one internal $\phi$ line. 
The initial Feynman integral is given by:
\begin{equation}
\Sigma_{\phi}^{(1L)}(p) = 4g^2 \int \frac{d^dk}{(2\pi)^d} \frac{1}{(k^2)^{\Delta_{\varphi}} \, ((p-k)^2)^{\Delta_{\phi}}} . \label{eq:sunset_integral}
\end{equation}
Combining the denominators via the Feynman parametrization and performing the momentum integration leads to the following expression:
\begin{equation}
\begin{aligned}
\Sigma_{\phi}^{(1L)}(p)
={}&
\frac{4g^2}{(4\pi)^{d/2}}
\frac{\Gamma(\Delta_{\phi}+\Delta_{\varphi}-d/2)}
{\Gamma(\Delta_{\phi})\Gamma(\Delta_{\varphi})}
\\
&\times
\int_0^1 dx\,
\frac{x^{\Delta_{\varphi}-1}(1-x)^{\Delta_{\phi}-1}}
{\bigl[x(1-x)p^2\bigr]^{
\Delta_{\phi}+\Delta_{\varphi}-d/2}} .
\end{aligned}
\label{eq:sunset_momentum_integrated}
\end{equation}
%\begin{widetext}
%\begin{equation}
%\Sigma_{\phi}^{(1L)}(p) = \frac{4g^2}{(4\pi)^{d/2}} \frac{\Gamma(\Delta_{\phi}+\Delta_{\varphi} - \frac{d}{2})}{\Gamma(\Delta_{\phi})\Gamma(\Delta_{\varphi})} \int_0^1 dx \frac{x^{\Delta_{\varphi}-1}(1-x)^{\Delta_{\phi}-1}}{[x(1-x)p^2]^{\Delta_{\phi}+\Delta_{\varphi}-d/2}} . \label{eq:sunset_momentum_integrated}
%\end{equation}
%\end{widetext}
Evaluating the remaining parameter integral through the Euler Beta function identity determines the coefficient function $B^{(1L)}(\Delta_{\phi}, \Delta_{\varphi})$ as a ratio of Euler Gamma functions:
\begin{equation}
B^{(1L)}(\Delta_{\phi}, \Delta_{\varphi}) = \frac{4\Gamma(\Delta_{\phi}+\Delta_{\varphi} - \frac{d}{2})\Gamma(\frac{d}{2}-\Delta_{\phi})\Gamma(\frac{d}{2}-\Delta_{\varphi})}{\Gamma(\Delta_{\phi})\Gamma(\Delta_{\varphi})\Gamma(d-\Delta_{\phi}-\Delta_{\varphi})} . \label{eq:B_final}
\end{equation}

\subsection{1-Loop Vertex Corrections}

The vertex correction is determined by the leading-order coefficient function $C(\Delta_\phi, \Delta_\varphi)$. 
At the one-loop level, evaluating the vertex correction at zero auxiliary-field external momentum yields
\begin{small}
\begin{equation}
\begin{split}
\Gamma_{g}^{(1L)} (p,0)&= 8g^2 \int \frac{d^dk}{(2\pi)^{d}} \frac{1}{((p-k)^2)^{2\Delta_\phi}(k^2)^{\Delta_\varphi}}\\
&=2\tilde{g}^2B^{(1L)}(2\Delta_\phi,\Delta_\varphi)(p^2)^{-\epsilon} . \label{eq:vertex_2l_integral}
\end{split}
\end{equation}
\end{small}
Comparison with Eq.~\eqref{eq:vertex_expansion} therefore identifies $C^{(1L)}(\Delta_\phi,\Delta_\varphi) =2B^{(1L)}(2\Delta_\phi,\Delta_\varphi)$.

\section{Functional RG Analysis and Fixed Point} \label{sec:wf_fixed_point}
The logarithmic corrections to the two quadratic kernels and to the local part of the Yukawa vertex are restricted to the leading order, omitting higher-loop contributions. 
Expanding the factors $(p^2)^{-\epsilon}$ for small $\epsilon$ yields the one-loop logarithmic derivatives:
\begin{align}
\frac{d\Pi_\varphi^{(1L)}}{dl}
&=
\tilde{g}^2
A^{(1L)}(\Delta_\phi)(2\epsilon)(p^2)^{\Delta_\varphi},
\label{eq:dPi_1loop_dl}
\\
\frac{d\Sigma_\phi^{(1L)}}{dl}
&=
\tilde{g}^2
B^{(1L)}(\Delta_\phi,\Delta_\varphi)(2\epsilon)(p^2)^{\Delta_\phi},
\label{eq:dSigma_1loop_dl}
\\
\frac{d\Gamma_g^{(1L)}}{dl}
&=
2\tilde{g}^2
B^{(1L)}(2\Delta_\phi,\Delta_\varphi)(2\epsilon).
\label{eq:dGamma_1loop_dl}
\end{align}
Here, $\epsilon=2\Delta_\phi+\Delta_\varphi-d/2$, and all couplings and exponents are evaluated at the running scale $l$. 
The RG convention is defined by
\[
p_{\rm old}=p_{\rm new}e^{-l},
\qquad
l=\ln\frac{\Lambda}{\mu}.
\]
The scale transformations for the fields and the coupling are given by
\[
\varphi_{\rm old}
=
\varphi_{\rm new}
\exp\Big[
\frac12\int_0^l dl'\,
\bigl(d+2\Delta_\varphi-\gamma_\varphi\bigr)
\Big],
\]
and
\[
\phi_{\rm old}
=
\phi_{\rm new}
\exp
\Big[
\frac12\int_0^l dl'\,
\bigl(d+2\Delta_\phi-\gamma_\phi\bigr)
\Big].
\]
Based on Eqs.~\eqref{eq:dPi_1loop_dl}--\eqref{eq:dGamma_1loop_dl}, the rescaled action after one infinitesimal RG step is expressed up to $O(dl^2)$ as
\begin{widetext}
\begin{align}
\mathcal S_{l+dl}
=&
\int_p
(1-\gamma_\phi dl)
\frac12
\phi(p)(p^2)^{\Delta_\phi}
\left[
1
-
2\epsilon
\frac{g(l)^2}{(4\pi)^{d/2}}
B^{(1L)}(\Delta_\phi,\Delta_\varphi)dl
\right]
\phi(-p)
\nonumber\\
&+
\int_p
(1-\gamma_\varphi dl)
\frac12
\varphi(p)(p^2)^{\Delta_\varphi}
\left[
1
-
2\epsilon
\frac{g(l)^2}{(4\pi)^{d/2}}
A^{(1L)}(\Delta_\phi)dl
\right]
\varphi(-p)
\nonumber\\
&+
g(l)
\left[
1+2\epsilon
\frac{g(l)^2}{(4\pi)^{d/2}}
B^{(1L)}(2\Delta_\phi,\Delta_\varphi)dl
\right]
\left[1+\left(\epsilon-\gamma_{\phi}-\frac{1}{2}\gamma_{\varphi}\right)dl\right]
\int_{p,k}
\phi(p-k)\varphi(k)\phi(-p).
\label{eq:action_after_infinitesimal_rg}
\end{align}
\end{widetext}
Here, $\int_p\equiv\int d^dp/(2\pi)^d$ and $\int_{p,k}\equiv\int d^dp\,d^dk/(2\pi)^{2d}$.

The anomalous dimensions are determined by requiring the quadratic kernels to retain their normalized form in Eq.~(\ref{eq:action_final}) after the RG step, yielding
\begin{equation}
\gamma_\phi
=
-2\epsilon
\frac{g^2}{(4\pi)^{d/2}}
B^{(1L)}(\Delta_\phi,\Delta_\varphi),
\label{eq:eta_phi_1loop}
\end{equation}
and
\begin{equation}
\gamma_\varphi
=
-2\epsilon
\frac{g^2}{(4\pi)^{d/2}}
A^{(1L)}(\Delta_\phi).
\label{eq:eta_varphi_1loop}
\end{equation}
The renormalized Yukawa coupling is identified from the coefficient of the local operator $\varphi\phi^2$. 
Expanding the final line of Eq.~\eqref{eq:action_after_infinitesimal_rg} to first order in $dl$ results in the relation:
\begin{small}
\begin{align}
g(l+dl)&=g(l) \nonumber \\
&+\left[\epsilon+2\epsilon\frac{g^2}{(4\pi)^{d/2}}B^{(1L)}(2\Delta_\phi,\Delta_\varphi)-\gamma_\phi-\frac12\gamma_\varphi\right]g(l)\,dl .
\end{align}
\end{small}
As a result, the one-loop beta function is given by
\begin{small}
\begin{equation}
\beta(g)\equiv\frac{dg}{dl}=\left[\epsilon+2\epsilon\frac{g^2}{(4\pi)^{d/2}}B^{(1L)}(2\Delta_\phi,\Delta_\varphi)-\gamma_\phi-\frac12\gamma_\varphi\right]g .
\label{eq:beta_form}
\end{equation}
\end{small}
Substituting Eqs.~\eqref{eq:eta_phi_1loop} and \eqref{eq:eta_varphi_1loop} into Eq.~\eqref{eq:beta_form} establishes the final expression for the beta function:
{\small
\begin{equation}
\begin{aligned}
\beta(g)
=&\,\epsilon g\Bigl[1+\tilde{g}^2\Bigl\{
2B^{(1L)}(2\Delta_\phi,\Delta_\varphi)
\\
&\quad
+2B^{(1L)}(\Delta_\phi,\Delta_\varphi)
+A^{(1L)}(\Delta_\phi)
\Bigr\}\Bigr].
\label{eq:beta_final}
\end{aligned}
\end{equation}
}

At a nontrivial fixed point where $g_*\neq0$, the following conditions are imposed:
\begin{equation}
\beta(g_*)=0,\qquad
\epsilon=2\Delta_\phi+\Delta_\varphi-\frac d2,\qquad |\epsilon|\ll1 .
\label{eq:fixed_point_condition}
\end{equation}

Defining the wave-function renormalization factors as $Z_{\varphi}=1-\tilde g_*^{\,2}A^{(1L)}(\Delta_\phi)$ and $Z_{\phi}=1-\tilde g_*^{\,2}B^{(1L)}(\Delta_\phi,\Delta_\varphi)$, the canonically normalized interaction vertex acquires the multiplicative factor $\sqrt{Z_{\varphi}}\,Z_{\phi}$. To implement amplitude matching for the interaction term $g\varphi\phi^2$, we impose the condition $\sqrt{Z_{\varphi}}\,Z_{\phi}=1$. Substituting the explicit one-loop expressions for the wave-function renormalization factors, the amplitude-matching condition is given by
\begin{equation}
\sqrt{1-\tilde g_*^2 A^{(1L)}(\Delta_\phi)}
\left[1-\tilde g_*^2 B^{(1L)}(\Delta_\phi,\Delta_\varphi)\right]=1 .
\label{eq:amplitude_matching}
\end{equation}
These three conditions determine the values of $(\Delta_\phi,\Delta_\varphi,\tilde g_*^2)$. 
The fixed-point equations admit multiple real solutions at a given value of $\epsilon$. Table~\ref{tab:d3_results} displays a representative subset of these solutions at $d=3$.

\begin{table*}[t!]
\centering
\caption{
Representative real solutions of the corrected fixed-point equations
at $d=3$. For each positive value of $\epsilon$, the first row is a
solution on the comparison branch selected for its relatively close
composite-sector agreement, while the second row, where displayed,
illustrates an alternative positive-coupling solution. Boldface denotes the solutions analyzed in detail in the text.
The benchmark values are derived from the conformal-bootstrap scaling
dimensions and critical exponents reported in Ref.~\cite{Kos_2016}.
Percentages in parentheses denote the absolute relative deviations
from the corresponding benchmark values.
}
\label{tab:d3_results}

\renewcommand{\arraystretch}{1.25}

\begin{tabular*}{\textwidth}
{@{\extracolsep{\fill}}cccccc@{}}
\hline\hline
$\epsilon$
& $\Delta_\phi$
& $\Delta_\varphi$
& $\widetilde g_*^{\,2}$
& $\Delta_{\phi^2}$
& $\nu$ \\
\hline

\multirow{2}{*}{$0.01$}
& $0.75374$ {\scriptsize $(23.23\%)$}
& $0.00253$
& $-25.96550$
& $1.49253$ {\scriptsize $(5.66\%)$}
& $0.66336$ {\scriptsize $(5.30\%)$} \\
&
$1.01340$ {\scriptsize $(3.21\%)$}
& $-0.51681$
& $0.08959$
& $0.97319$ {\scriptsize $(31.11\%)$}
& $0.49339$ {\scriptsize $(21.68\%)$} \\
\hline

\multirow{2}{*}{$0.10$}
& $0.78647$ {\scriptsize $(19.90\%)$}
& $0.02707$
& $-2.21296$
& $1.42707$ {\scriptsize $(1.02\%)$}
& $0.63575$ {\scriptsize $(0.92\%)$} \\
&
$1.13151$ {\scriptsize $(15.24\%)$}
& $-0.66301$
& $0.07563$
& $0.73699$ {\scriptsize $(47.83\%)$}
& $0.44189$ {\scriptsize $(29.86\%)$} \\
\hline

\multirow{2}{*}{$0.30$}
& $\mathbf{0.86266}$ {\scriptsize $\mathbf{(12.14\%)}$}
& $\mathbf{0.07467}$
& $\mathbf{-0.53955}$
& $\mathbf{1.27467}$ {\scriptsize $\mathbf{(9.77\%)}$}
& $\mathbf{0.57960}$ {\scriptsize $\mathbf{(8.00\%)}$} \\
&
$1.35867$ {\scriptsize $(38.38\%)$}
& $-0.91735$
& $0.02132$
& $0.28265$ {\scriptsize $(79.99\%)$}
& $0.36801$ {\scriptsize $(41.58\%)$} \\
\hline

$0.50$
& $0.95790$ {\scriptsize $(2.44\%)$}
& $0.08420$
& $-0.28295$
& $1.08420$ {\scriptsize $(23.25\%)$}
& $0.52197$ {\scriptsize $(17.14\%)$} \\
\hline

$-0.198$
& $\mathbf{0.97714}$ {\scriptsize $\mathbf{(0.48\%)}$}
& $\mathbf{-0.65260}$
& $\mathbf{0.01864}$
& $\mathbf{1.04573}$ {\scriptsize $\mathbf{(25.97\%)}$}
& $\mathbf{0.51170}$ {\scriptsize $\mathbf{(18.77\%)}$} \\
\hline

\multirow{3}{*}{$-0.248$}
& $0.76575$ {\scriptsize $(22.01\%)$}
& $-0.27950$
& $0.00298$
& $1.46850$ {\scriptsize $(3.96\%)$}
& $0.65295$ {\scriptsize $(3.65\%)$} \\
&
$\mathbf{0.80473}$ {\scriptsize $\mathbf{(18.04\%)}$}
& $\mathbf{-0.35746}$
& $\mathbf{0.00840}$
& $\mathbf{1.39054}$ {\scriptsize $\mathbf{(1.56\%)}$}
& $\mathbf{0.62133}$ {\scriptsize $\mathbf{(1.37\%)}$} \\
&
$0.85470$ {\scriptsize $(12.95\%)$}
& $-0.45740$
& $0.01325$
& $1.29060$ {\scriptsize $(8.64\%)$}
& $0.58500$ {\scriptsize $(7.14\%)$} \\
\hline

Benchmark
& $0.98185$
& ---
& ---
& $1.41263$
& $0.62997$ \\
\hline\hline
\end{tabular*}
\end{table*}

For $d=3$, high-precision Monte Carlo simulations provide, consistent estimates of the Ising critical exponents \cite{Hasenbusch_2010,Ferrenberg_2018}.
For the quantitative comparisons below, we use the conformal-bootstrap benchmark values $\eta_{\phi}^*=0.0362978(20)$, $\Delta_{\phi^2}^*=1.412625(10)$, and $\nu^*=0.629971(4)$ \cite{Kos_2016}.
The corresponding benchmark value of $\Delta_\phi$ is obtained from
\begin{equation}
\eta_\phi = 2 - 2\Delta_\phi,
\end{equation}
which gives $\Delta_\phi^*\approx0.98185$. 
The remaining critical exponents are determined through
\begin{equation}
\nu = \frac{1}{d-\Delta_{\phi^2}},
\end{equation}
and
\begin{equation}
\Delta_{\phi^2}
= d-[g]-[\varphi(x)]
= \frac{d}{2}+\Delta_\varphi-[g].
\end{equation}

Table~\ref{tab:d3_results} lists multiple real solutions of the
fixed-point equations at $d=3$. We first focus on the solution at
$\epsilon=-0.198$, which gives the closest agreement in the
fundamental-field sector among the tabulated solutions while retaining
a small positive fixed-point coupling. We select this solution as the
primary representative:
\begin{equation}
    \Delta_\phi \approx 0.97714, \qquad
    \Delta_{\phi^2} \approx 1.04573,
\end{equation}
\begin{equation}
    \eta_\phi \approx 0.04572, \qquad
    \nu \approx 0.51170,
\end{equation}
with $\widetilde{g}_{*}^{\,2}\approx0.01864$.
Compared with the conformal-bootstrap benchmark values
$\Delta_\phi^*\approx0.9818511$,
$\Delta_{\phi^2}^*\approx1.412625$, and
$\nu^*\approx0.629971$ \cite{Kos_2016}, the relative deviations are
approximately $0.48\%$, $25.97\%$, and $18.77\%$, respectively.
Thus, this solution accurately reproduces the fundamental-field
scaling, although the agreement is substantially less precise in the
composite and correlation-length sectors.

For comparison, the nearby solution at $\epsilon=-0.248$ exhibits a
complementary pattern of agreement. At this value of $\epsilon$, three
neighboring solutions have positive
$\widetilde{g}_{*}^{\,2}$, and the middle solution provides the closest
agreement in the composite-field sector:
\begin{equation}
    \Delta_\phi \approx 0.80473, \qquad
    \Delta_{\phi^2} \approx 1.39054,
\end{equation}
\begin{equation}
    \eta_\phi \approx 0.39054, \qquad
    \nu \approx 0.62133,
\end{equation}
with $\widetilde{g}_{*}^{\,2}\approx0.00840$.
The relative deviations in $\Delta_\phi$, $\Delta_{\phi^2}$, and
$\nu$ are approximately $18.04\%$, $1.56\%$, and $1.37\%$,
respectively. In contrast to the primary $\epsilon=-0.198$ solution,
this branch reproduces the composite-operator dimension and the
correlation-length exponent within approximately $2\%$ of the
benchmark values, while showing substantially lower accuracy in the
fundamental-field sector. Together, the two solutions illustrate the
sector-dependent accuracy of the one-loop closure.

As a further comparison, we also highlight the solution at the
representative positive value $\epsilon=0.30$. Among the tabulated
positive-$\epsilon$ solutions, this branch exhibits the most even
agreement between the fundamental- and composite-field sectors:
\begin{equation}
    \Delta_\phi \approx 0.86266, \qquad
    \Delta_{\phi^2} \approx 1.27467,
\end{equation}
\begin{equation}
    \eta_\phi \approx 0.27467, \qquad
    \nu \approx 0.57960,
\end{equation}
with $\widetilde{g}_{*}^{\,2}\approx-0.53955$.
The relative deviations in $\Delta_\phi$, $\Delta_{\phi^2}$, and
$\nu$ are approximately $12.14\%$, $9.77\%$, and $8.00\%$,
respectively. Compared with the two negative-$\epsilon$ branches
discussed above, the agreement is more evenly distributed across the
fundamental, composite, and correlation-length sectors, although it
remains moderate overall and is accompanied by a negative value of
$\widetilde{g}_{*}^{\,2}$.

A notable contrast between the two representative solutions concerns the sign and magnitude of the fixed-point coupling. For $\epsilon=0.30$, the comparison solution has $\widetilde{g}_{*}^{\,2}\approx-0.53955$, whereas the representative solution at $\epsilon=-0.248$ has the small positive value $\widetilde{g}_{*}^{\,2}\approx0.00840$.
In conventional local quantum field theories, a negative coupling typically indicates a violation of unitarity or an unphysical fixed point. 
Within the present framework, however, this property requires careful interpretation; because the effective action ansatz incorporates non-local kinetic kernels with unconstrained scaling exponents, standard criteria for local unitarity are not directly applicable. 
Nevertheless, given that the critical exponents derived from this one-loop truncation show numerical deviations from high-precision conformal bootstrap benchmarks, this negative value is likely a one-loop artifact. 
Higher-loop corrections, such as a two-loop skeleton diagrammatic evaluation, remain necessary to clarify whether a physical fixed point with a real, positive coupling emerges when vertex fluctuations are more comprehensively incorporated.

\section{Conclusion}

In summary, this work investigates the three-dimensional Wilson-Fisher fixed point within an ansatz-driven Renormalization Group (RG) scheme by treating the scaling dimensions as unconstrained variables. 
The two-variable self-consistent scheme determines the RG flows of the couplings and the scaling exponents at the leading one-loop order. 
While the resulting critical exponents show numerical deviations from high-precision Monte Carlo and conformal bootstrap benchmarks, the framework provides a self-consistent field-theoretic scheme based on a nonlocal effective action to evaluate the scaling data without fixing the intermediate scaling dimensions.

To contextualize this framework, the results may be compared with conventional perturbative methodologies, such as the $\epsilon$-expansion ($d = 4 - \epsilon$). 
The perturbative renormalization-group functions underlying the $\epsilon$-expansion have been computed through six loops \cite{Kompaniets_2017} and subsequently extended to seven loops \cite{Schnetz_2018}. 
The associated $\epsilon$-expansions of the critical exponents are factorially divergent asymptotic series and therefore require resummation for quantitative estimates in three dimensions \cite{Kompaniets_2017}.
Quantitative estimates in three dimensions are then obtained using resummation procedures, such as Borel resummation combined with conformal mapping \cite{Kompaniets_2017} or hypergeometric--Meijer resummation \cite{Shalaby_2020,Shalaby_2021}. 
In contrast, the nonlocal effective action ansatz introduces an alternative structure where the scaling dimensions of the primary and auxiliary fields act as independent variables. 
The current leading-order calculation establishes the baseline scaling equations, avoiding the need for multi-loop series resummation at this initial stage. 
Further extensions are required to improve the quantitative accuracy and address the numerical discrepancies observed at the one-loop level.

Beyond these quantitative improvements, the present unfreezing of the auxiliary-field kernel provides a robust starting point for accessing the complete kinematic dependence of the theory. While the current closure already confronts the fractional powers of the non-local kernels---requiring specific branch and phase conventions under analytic continuation to evaluate the loops in Sections III A, B, and C---these evaluations remain projected onto specific kinematic channels to extract the quadratic kernels and local Yukawa-vertex flow. 

Precisely because of these underlying fractional structures, any comprehensive extension to the full four-point sectors, as probed by
\begin{equation}
\bigl\langle\phi\phi\phi\phi\bigr\rangle, \qquad \bigl\langle\phi\phi\varphi\varphi\bigr\rangle,
\end{equation}
inherently demands resolving their non-trivial cross-ratio dependence, while their conformal-block decompositions would encode and constrain the associated operator product expansion (OPE) data \cite{SimmonsDuffin_2017}. At the four-point level, the resulting correlators must satisfy crossing compatibility among different OPE channels and remain single-valued in Euclidean kinematics, even though individual conformal-block contributions may carry non-trivial monodromy \cite{SimmonsDuffin_2014, Karateev_2019}. These analytic and higher-point consistency conditions will provide a stringent test of whether the candidate branches obtained here admit a fully consistent completion as a critical theory.

\begin{acknowledgments}
K.-S. K. was supported by the Ministry of Education, Science, and Technology (Grant No. RS-2024-00337134) of the National Research Foundation of Korea (NRF). J. M. B. acknowledges support from the National Research Foundation of Korea (NRF) (No. RS-2022-NR072365).
\end{acknowledgments}


\begin{thebibliography}{99}

\bibitem{Wilson_Fisher_1972}
K. G. Wilson and M. E. Fisher,
\textit{Critical Exponents in 3.99 Dimensions},
Phys. Rev. Lett. \textbf{28}, 240 (1972).

\bibitem{Wilson_Kogut_1974}
K. G. Wilson and J. Kogut,
\textit{The renormalization group and the $\epsilon$ expansion},
Phys. Rep. \textbf{12}, 75 (1974).

\bibitem{Sachdev_2011}
S. Sachdev,
\textit{Quantum Phase Transitions}, 2nd ed.
(Cambridge University Press, Cambridge, 2011).

\bibitem{LeGuillou_1977}
J. C. Le Guillou and J. Zinn-Justin,
\textit{Critical Exponents for the $n$-Vector Model in Three
Dimensions from Field Theory},
Phys. Rev. Lett. \textbf{39}, 95 (1977).

\bibitem{Pelissetto_Vicari_2002}
A. Pelissetto and E. Vicari,
\textit{Critical phenomena and renormalization-group theory},
Phys. Rep. \textbf{368}, 549 (2002).

\bibitem{Moshe_ZinnJustin_2003}
M. Moshe and J. Zinn-Justin,
\textit{Quantum field theory in the large $N$ limit: A review},
Phys. Rep. \textbf{385}, 69 (2003).

\bibitem{Berges_2002}
J. Berges, N. Tetradis, and C. Wetterich,
\textit{Non-perturbative renormalization flow in quantum field theory
and statistical physics},
Phys. Rep. \textbf{363}, 223 (2002).

\bibitem{Rosten_2012}
O. J. Rosten,
\textit{Fundamentals of the Exact Renormalization Group},
Phys. Rep. \textbf{511}, 177 (2012).

\bibitem{Hasenbusch_2010}
M. Hasenbusch,
\textit{Finite size scaling study of lattice models in the
three-dimensional Ising universality class},
Phys. Rev. B \textbf{82}, 174433 (2010).

\bibitem{Ferrenberg_2018}
A. M. Ferrenberg, J. Xu, and D. P. Landau,
\textit{Pushing the limits of Monte Carlo simulations for the
three-dimensional Ising model},
Phys. Rev. E \textbf{97}, 043301 (2018).

\bibitem{ElShowk_2012}
S. El-Showk, M. F. Paulos, D. Poland, S. Rychkov,
D. Simmons-Duffin, and A. Vichi,
\textit{Solving the 3D Ising model with the conformal bootstrap},
Phys. Rev. D \textbf{86}, 025022 (2012).

\bibitem{Kos_2016}
F. Kos, D. Poland, D. Simmons-Duffin, and A. Vichi,
\textit{Precision islands in the Ising and $O(N)$ models},
J. High Energy Phys. \textbf{08}, 036 (2016).

\bibitem{Fei_Giombi_Klebanov_2014}
L. Fei, S. Giombi, and I. R. Klebanov,
\textit{Critical $O(N)$ models in $6-\epsilon$ dimensions},
Phys. Rev. D \textbf{90}, 025018 (2014).

%\bibitem{Kos_2014}
%F. Kos, D. Poland, and D. Simmons-Duffin,
%\textit{Bootstrapping the $O(N)$ vector models},
%J. High Energy Phys. \textbf{06}, 091 (2014).

%\bibitem{Bervillier_2007}
%C. Bervillier, A. J{\"u}ttner, and D. F. Litim,
%\textit{High-accuracy scaling exponents in the local potential
%approximation},
%Nucl. Phys. B \textbf{783}, 213 (2007).

\bibitem{Kompaniets_2017}
M. V. Kompaniets and E. Panzer,
\textit{Minimally subtracted six loop renormalization of
$O(n)$-symmetric $\phi^4$ theory and critical exponents},
Phys. Rev. D \textbf{96}, 036016 (2017).

\bibitem{Schnetz_2018}
O. Schnetz,
\textit{Numbers and functions in quantum field theory},
Phys. Rev. D \textbf{97}, 085018 (2018).

%\bibitem{Kazakov_1979}
%D. I. Kazakov, O. V. Tarasov, and D. V. Shirkov,
%\textit{Analytic continuation of the results of perturbation theory
%for the model $g\phi^4$ to the region $g\gtrsim 1$},
%Theor. Math. Phys. \textbf{38}, 9 (1979).

\bibitem{Shalaby_2020}
A. M. Shalaby,
\textit{Precise critical exponents of the $O(N)$-symmetric quantum
field model using hypergeometric-Meijer resummation},
Phys. Rev. D \textbf{101}, 105006 (2020).

\bibitem{Shalaby_2021}
A. M. Shalaby,
\textit{Critical exponents of the $O(N)$-symmetric $\phi^4$ model
from the $\epsilon^7$ hypergeometric-Meijer resummation},
Eur. Phys. J. C \textbf{81}, 87 (2021).

\bibitem{SimmonsDuffin_2017} 
D. Simmons-Duffin, 
\textit{The conformal bootstrap}, in \textit{New Frontiers in Fields and Strings} (World Scientific, Singapore, 2017), pp.~1--74. 

\bibitem{SimmonsDuffin_2014} 
D. Simmons-Duffin, 
\textit{Projectors, shadows, and conformal blocks}, 
J. High Energy Phys. \textbf{04}, 146 (2014). 

\bibitem{Karateev_2019} 
D. Karateev, P. Kravchuk, and D. Simmons-Duffin, 
\textit{Harmonic analysis and mean field theory}, 
J. High Energy Phys. \textbf{10}, 217 (2019).

\end{thebibliography}
\end{document}